%% file: main.tex
\documentclass[conference,preprint]{vgtc}             %

\ifpdf%
  \pdfoutput=1\relax                   %
  \usepackage{graphicx}                %
  \DeclareGraphicsExtensions{.pdf,.png,.jpg,.jpeg} %
\else%
  \usepackage{graphicx}                %
  \DeclareGraphicsExtensions{.eps}     %
\fi%

\graphicspath{{figures/}{pictures/}{images/}{./}} %

\usepackage{microtype}                 %
\PassOptionsToPackage{warn}{textcomp}  %
\usepackage{textcomp}                  %
\usepackage{mathptmx}                  %
\usepackage{times}                     %
\usepackage{cite}                      %
\usepackage{tabu}                      %
\usepackage{booktabs}                  %
\usepackage{soul,color}

\onlineid{0}

\vgtccategory{Research}
\vgtcpapertype{please specify}

\title{Periphery Plots for Contextualizing Heterogeneous Time-Based Charts}

\author{Bryce Morrow, Trevor Manz, Arlene E. Chung, Nils Gehlenborg, David Gotz}
\authorfooter{
\item
 Bryce Morrow is with the University of North Carolina at Chapel Hill. E-mail: bam4564@live.unc.edu.
\item
 Trevor Manz is with Harvard Medical School. Email: trevor_manz@hms.harvard.edu.
\item
 Arlene E. Chung is with the University of North Carolina School of Medicine. E-mail: arlene_chung@med.unc.edu.
\item
 Nils Gehlenborg is with Harvard Medical School. Email: nils@hms.harvard.edu.
\item
 David Gotz is with the University of North Carolina at Chapel Hill. Email: gotz@unc.edu.
}

\shortauthortitle{Morrow \MakeLowercase{\textit{et al.}}: Periphery Plots}

\abstract{
\vspace{-0.05cm}
Patterns in temporal data can often be found across different scales, such as days, weeks, and months, making effective visualization of time-based data challenging. Here we propose a new approach for providing focus and context in time-based charts to enable interpretation of patterns across time scales. Our approach employs a focus zone with a time and a second axis, that can either represent quantities or categories, as well as a set of adjacent periphery plots that can aggregate data along the time, value, or both dimensions. We present a framework for periphery plots and describe two use cases that demonstrate the utility of our approach.} %


\teaser{
  \centering
  \vspace{-0.1cm}
  \includegraphics[width=0.968\linewidth]{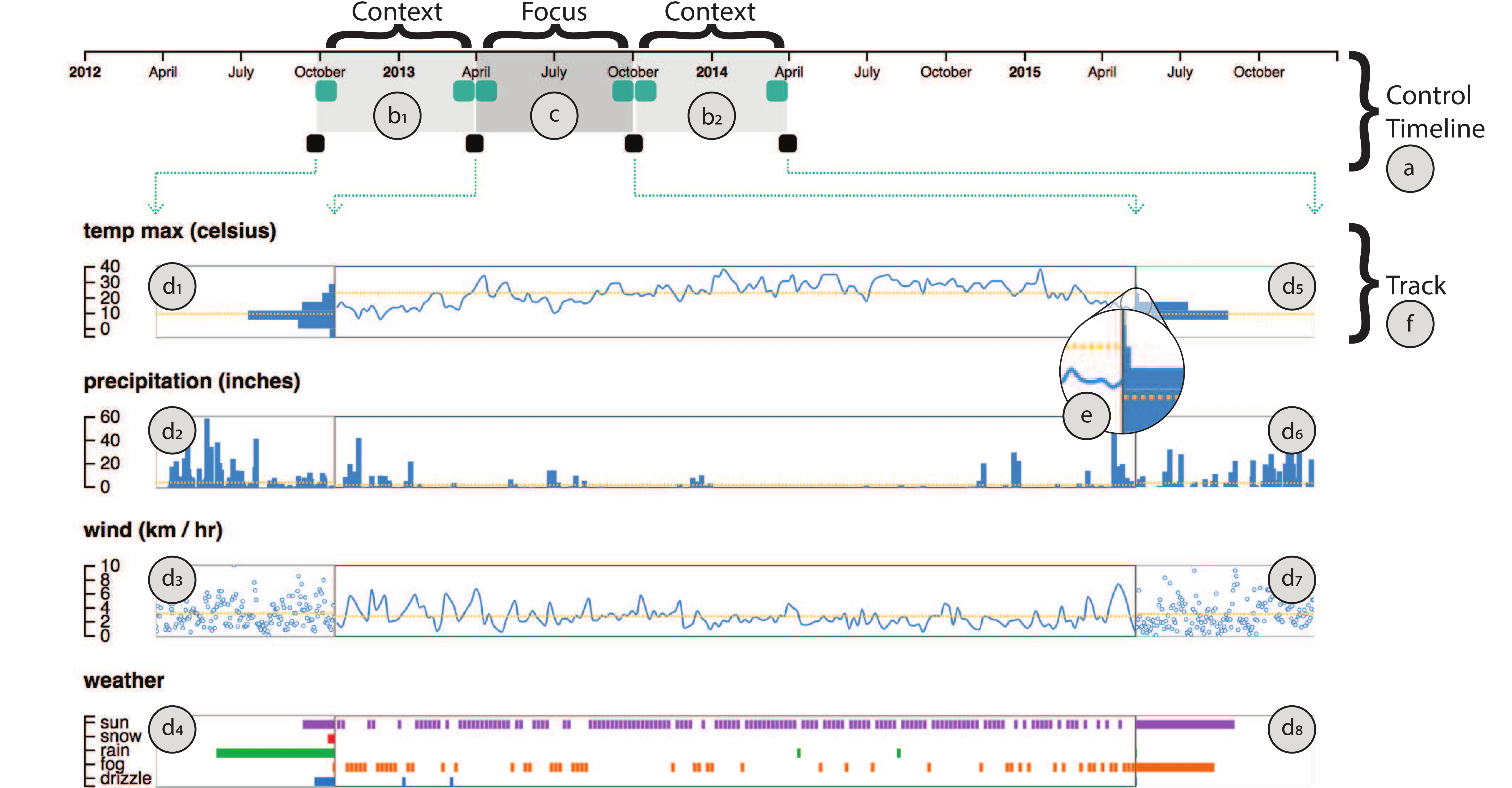}
  \vspace{-0.2cm}
  \caption{Periphery plots used within a visualization of multi-variate climate data. The (a) control timeline includes interactive controls to define (b$_1$, b$_2$) context and (c) the focus zone. This example includes four tracks (f), each with (d$_1$-d$_8$) two periphery plots. (e) Annotations support quick comparisons, such as the seasonal variation between average maximum temperatures in summer months (the focus region) and the preceding and subsequent winter periods (d$_1$, d$_5$).}
	\label{fig:teaser}
}

\vgtcinsertpkg

\begin{document}
    \vspace{-0.15cm}
\input{sections/intro.tex}
\input{sections/related.tex}

\input{sections/design.tex}

\input{sections/prototype.tex}
\input{sections/conclusion.tex}

\acknowledgments{
\vspace{-0.05cm}
This research was supported by the National Institutes of Health under award number 5R01EB025024. Additional support was provided by the National Science Foundation via Grant \#1704018.
}

\bibliographystyle{abbrv-doi-narrow}

\clearpage
\bibliography{gotz.bib,VAST2019.bib}
\end{document}

%% file: sections/intro.tex
\vspace{-0.05cm}
\firstsection{Introduction}

\maketitle

\vspace{-0.05cm}
Temporal data is found frequently in various application domains and comes in many forms. Data can range from sparse point-based and interval events (e.g.,~procedures and medication changes) to densely sampled time series with repeated scalar measurements (e.g.,~longitudinal climate measurements). Often, data have combinations of variables and data types (e.g.,~daily stock prices and trading volumes along with the dates of key news events).

Based on these temporal data sets, users must often explore data and answer complex temporal questions. For example, consider a clinician exploring data gathered from mobile health (mHealth) devices (e.g.,~fitness trackers and mobile phones) along with patient-reported data (e.g.,~diet and patient-reported outcomes about health status) and clinical data (e.g.,~medications and diagnoses). To make sense out of this kind of data, one must: (1) view data across a range of time scales (e.g.,~daily exercise fluctuations vs. monthly patient-reported outcomes vs. semi-annual doctor visits), (2) correlate across variables (e.g.,~does a change in medication correlate with changes in patient-reported outcomes?), and (3) compare across time intervals (e.g.,~has a patient's current condition improved compared to historical data?).

Given the ubiquity of temporal data, a wide variety of visual analytics techniques have been proposed for representing time-based events. This includes multi-variate techniques as well as a range of different \textit{focus + context} methods. However, these techniques often focus on homogeneous types of data (e.g.,~scalar valued time series) or have other constraints that limit their use (see Section~\ref{sec:related}). 
This paper presents \emph{periphery plots}, a technique for augmenting time-aligned temporal charts with a set of horizontally aligned peripheral views that provide multi-resolution contexts before and after a focal temporal period.
This approach is applicable to heterogeneous data types 
and varied time samplings (regular vs. irregular; hourly vs. annually). Moreover, it is capable of representing a variety of previously proposed \textit{focus + context} methods, while simultaneously supporting new contextualization capabilities.
More specifically, this research has three key contributions:

\vspace{-0.18cm}
\begin{itemize}
    \item {\bf A Framework for Periphery Plots.} The framework includes key concepts of focus and context zones. It also includes an axis preservation design space for plots that allow visualizations with varying levels of summarization.
    
\vspace{-0.18cm}

    \item {\bf A Periphery Plot Toolkit.} The framework is implemented within an open-source toolkit that includes a collection of specific view types within a general purpose architecture.%
    
\vspace{-0.18cm}
    \item {\bf Example Applications.} Two use cases (longitudinal climate data analysis and pattern finding within patient-generated health data) demonstrate the utility of periphery plots.  
\end{itemize}

\vspace{-0.12cm}

%% file: sections/related.tex
\vspace{-0.15cm}
\section{Related Work}
\vspace{-0.10cm}
\label{sec:related}

This section provides a brief review of the most relevant themes from prior research on temporal data visualization.

\vspace{-0.08cm}
\subsection{Visual Representations of Temporal Data}
\vspace{-0.05cm}

Time series data sets are ubiquitous and are most commonly represented as line charts. Prior research has mainly focused on developing alternative visual encodings to convey equal information while using less space, such as horizon charts \cite{few_time_2008} and braided graphs \cite{javed_graphical_2010}. For horizon charts, interpretability can be preserved and improved while reducing space usage by optimizing chart parameters (number of bands, height of chart) \cite{heer_sizing_2009}.
Another approach is to create complex visual encodings with high information density. For example, both ThemeRiver\cite{havre_themeriver:_2000} and RankExplorer \cite{shi_rankexplorer:_2012} utilize stacked graphs as a base chart to display categorical partitions of multi-variate time series data, while the latter layers additional glyphs to represent changes in information density between categories and time points. 
Alternatively, Chronodes condenses multiple features into a single event timeline to highlight frequent event sub-sequences, where glyphs encode unique health related features \cite{polack_chronodes:_2018}. 
The periphery plots framework is compatible with a wide range of these aforementioned techniques because it supports the use of both common and bespoke chart types. Moreover, it can dynamically transition between types in response to changes in data density.

\vspace{-0.08cm}
\subsection{Detail and Context Interfaces}
\vspace{-0.05cm}
Exploratory analysis often requires users to interact with data at a level of detail that cannot be displayed on a single screen. While interactive methods such as \textit{zooming \& panning} offer the ability to view details otherwise hidden due to summarization, these interactions alone may obscure context and hinder interpretation due to narrow fields-of-view \cite{borland_contextual_2018}. A variety of techniques
have been developed to address this challenge, including \textit{overview~+~detail} and \textit{focus~+~context} approaches. The former employs spatial separation to partition global and detailed information (e.g., Stack Zooming \cite{javed_stack_2013}) while the latter combines these displays in a single continuous view (e.g., PerspectiveWall \cite{mackinlay_perspective_1991} or fish-eye lenses \cite{cockburn_review_2008,tory_human_2004}). 

Complementary approaches which incorporate these schemes with interactive techniques have yielded powerful dynamic tools for visual exploration, particularly for temporal data.  
SignalLens utilizes \textit{focus + context} with panning to enable rapid navigation of time series data, but users are unable to define the initial focus from a global overview \cite{kincaid_signallens:_2010}. In contrast, MultiStream \cite{cuenca_multistream:_2018} combines \textit{overview + detail} and \textit{focus + context}, linking interactive brush selections from a global overview to contextualize the focus of a fisheye distortion in a more detailed multi-resolution view. Our proposed periphery plot framework uses a similar combined \textit{overview + detail} and \textit{focus + context} approach. However, it focuses on views of context zones (rather than the focus) and includes a design space for various levels of contextual summarization.

\vspace{-0.08cm}
\subsection{Design Space for Multi-variate Time-Based Charts}
\vspace{-0.05cm}

KronoMiner \cite{zhao_kronominer:_2011} presents a taxonomy for the design space of layouts for visualizing multi-variate, time-based data. It includes stacking (i.e., multiple views are stacked on an axis to create global time alignment) and overplotting (i.e., multiple visual encodings are layered in a single view).  
Our framework supports both techniques and adds the ability to utilize multiple non-overlapping encodings (one for each zone) in a single stacked track.

%% file: sections/design.tex
\vspace{-0.05cm}
\section{Periphery Plots Framework}
\vspace{-0.05cm}

\input{sections/design_part_1.tex}

\input{sections/design_part_2.tex}

\input{sections/design_part_3.tex}

%% file: sections/design_part_1.tex
The periphery plots framework provides multi-scale contextualization capabilities for viewing heterogeneous time-based data, and includes several core concepts for how periphery plots connect with each other and other visualization elements. The framework also defines a two-dimensional design space for periphery plots themselves and a set of common user interactions.

\vspace{-0.05cm}
\subsection{Concepts}
\vspace{-0.05cm}

Periphery plots are defined by the relationship between temporally linked \emph{focus} and \emph{context zones}, which represent user-specified intervals along a dataset's time axis. These zones are used to determine the intervals of data to be visualized within time-based \emph{tracks}. Each track includes a set of \emph{periphery plots} and a \emph{focus plot} with one plot for each zone. The zones can be manipulated via interactions with the plots on individual tracks or through a global \emph{control timeline}.

\textbf{Focus and Context Zones.} Zones are the primary concept that underpins the framework for periphery plots. %
For a given temporal data visualization, periphery plots define a single focus zone and a set of context zones. Each zone is an interval of time defined by start and end points that are linked together in a non-overlapping, contiguous sequential ordering. Figure~\ref{fig:teaser}(b$_1$,c,b$_2$) shows an example that has two context zones, one to each side of the focus zone. The framework allows multiple context zones on either side of the focus as demonstrated in the accompanying video figure.

The focus zone corresponds to the time period of primary interest. The visualization used for the focus period is designed to show data at high resolution for detailed analysis (e.g., a line chart, a horizon graph, or an event timeline) with the visual design dependent, in part, on the specific data type and application. The focus zone is equivalent to the interval of temporal data displayed in a more typical visualization that does not include periphery plots.

In contrast, context zones are defined relative to the focus zone as either pre- or post-focus intervals of time.  Context zones are not generally the user's primary focus of attention.  Instead, they represent neighboring periods of time to the focus which provide context to users during interpretation of data found within the focus zone.

While it is common to define two context zones (one pre- and one post-focus), periphery plots allow multiple context zones to enable multi-scale contextualization (e.g, six context zones can be used to contextualize a focus time interval with a week, month, or year before and afterwards). Thus, an arbitrary set of $n$ context zones are defined relative to a single focus zone. 

Each context zone is visualized using periphery plots. As the screen space allocated to a periphery plot is typically more constrained than the focus, data summarization is a key element in designing effective periphery plots. This motivates the periphery plot design space outlined in Section~\ref{sec:design_space}.

\textbf{Tracks.} A track is a container that horizontally aligns a set of visualizations that correspond, typically, to a single variable. As shown in Figure~\ref{fig:teaser}(f), each track contains a periphery plot for each context zone and a single focus visualization corresponding to the focus zone. The visualizations within a track are arranged contiguously from left to right, ordered by the sequence of time intervals for the corresponding context and focus zones.

Multiple tracks can be used to visualize multiple variables simultaneously. In this case, all tracks are defined using the same set of context and focus zones. In addition, the tracks are stacked vertically with boundaries between zones aligned as shown in Figures~\ref{fig:teaser} and~\ref{fig:health}. This facilitates comparison across tracks by lining up data from the same time interval in vertical columns. The visual encodings for each track can be configured independently, supporting the visualization of heterogeneous data types.

\textbf{Control timeline.} The specification of context and focus zones is supported via a single control timeline. This mechanism includes a set of multiple-linked brushes that map one-to-one to the individual zones. Any changes to the focus or context zones made through the control timeline are tightly coordinated with the tracks to support a range of interactions as described in Section~\ref{sec:interactions}.

%% file: sections/design_part_2.tex
\subsection{Periphery Plot Design Space}
\label{sec:design_space}

Each context zone is visualized using one periphery plot per track. As previously described, periphery plots are typically space constrained which makes summarization a key design goal. We characterize the design space for periphery plots in terms of an \textbf{Axis Preservation Model} that includes a two-dimensional taxonomy for visual encodings of temporal data. The model categorizes encodings based on whether or not they map the \emph{time domain} and/or \emph{value domain} from data space to the $x$ and $y$ coordinates of screen space, respectively. This model is used for both continuous and discrete values (e.g., ordinal values or nominal categories).  

\begin{figure}[t]
\centering
\includegraphics[width=2.55in]{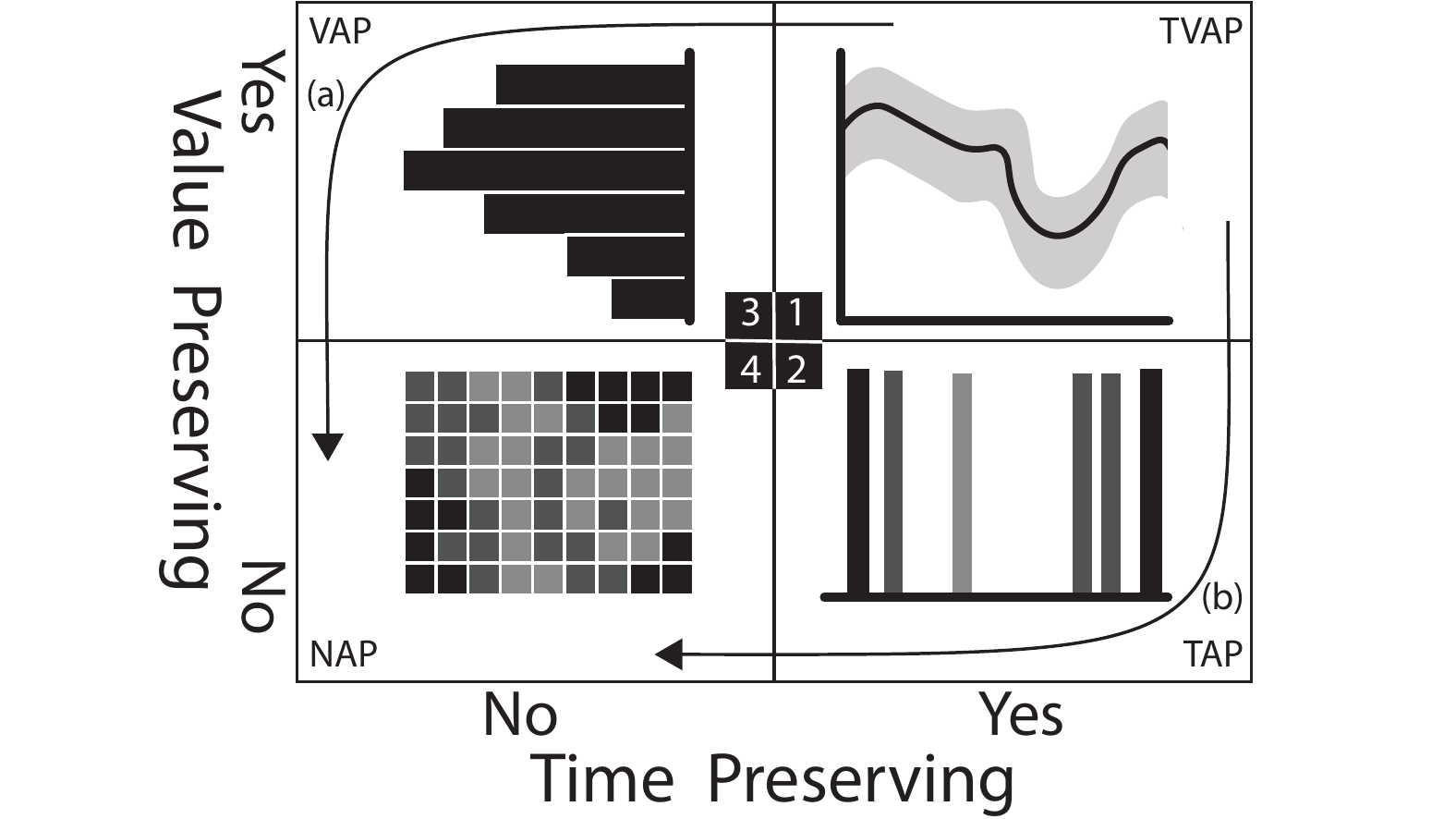}
\vspace{-0.3cm}
\caption{Periphery plot types can be characterized by their preservation of the time and value axes.  
Summarization can prioritize (a) value preservation or (b) time preservation.}
\vspace{-0.25cm}
\label{fig:2x2}
\end{figure}

This model leads to four types of periphery plots as illustrated in Figure~\ref{fig:2x2}:
time-value-axis preserving (TVAP),
time-axis preserving (TAP), 
value-axis preserving (VAP), and 
no-axis preserving (NAP).

TVAP plots, such as the bar charts in Figure~\ref{fig:teaser}(d$_2$) or dot plots in Figure~\ref{fig:teaser}(d$_3$), present the most detailed information. As data gets dense, aggregation by value or time can lead to more summarized views. For example, VAP plots aggregate over time but preserve values as in the histograms in Figure~\ref{fig:teaser}(d$_1$,d$_4$). TAP plots, in contrast, aggregate by value but preserve time. For example, a vertical band chart that shows the counts of observations per time period without distinguishing between values of observations as depicted symbolically in Figure~\ref{fig:2x2}(2). NAP plots adopt designs that preserve neither time nor value, such as grid-based or pixel-based visualizations.

We note that the axis preservation model applies to a variety of time- and value-based transformations. For example, a moving average envelope (e.g., Figure~\ref{fig:2x2}(1)) that aggregates values within a sliding time window would be classified as TVAP because it maintains time as a dimension despite the windowing transformation.

\textbf{Summarization.} Axis preservation can be viewed as an approach to summarization through relaxed constraints on visual encoding. Preserving both time and value places constraints on visual design that can make it difficult for periphery plots to effectively display the large scale volumes of data that often fall within a context zone. This problem is exacerbated by the limited screen space typically allocated to periphery plots. %

Sacrificing an axis allows for more flexibility in visual encoding. This can make visualizations less informative in some ways (the omitted axis), but enables new types of more space-efficient visual representations. This observation imposes an ordering on plot types with increasing levels of summarization: TVAP $\rightarrow$ (VAP,TAP) $\rightarrow$ NAP. Transitions between periphery plot types are a natural approach to handling changes in data volume within a context zone, and two summarization paths fit naturally within the axis preservation model as shown by the arrows in Figure~\ref{fig:2x2}(a,b). Moreover, such transitions can be automated as discussed in Section~\ref{sec:health}.

\textbf{Annotations.} To facilitate comparisons between zones within a single track, our design introduces \emph{annotations}. These are layered graphics that allow for comparisons of derived measures between plots, including plots that implement different visual encodings. Many common types of derived measurements can be compared via annotations, including averages, confidence intervals, and quantiles. 

The \textit{temp max} track in Figure \ref{fig:teaser} demonstrates the utility of average line annotations for identifying differences in values in the focus area and the context zones. As the orange dotted line annotations in Figure~\ref{fig:teaser}(e) show, the value within the focus zone is higher than in the subsequent time period. As demonstrated in the accompanying video, further insights can be obtained from dynamic changes in annotations in response to user interactions. This is because those interactions can result in different intervals of data being displayed in both focus and context zones.

%% file: sections/design_part_3.tex
\subsection{Interactions}
\label{sec:interactions}

Periphery plots can be informative even for static visualizations. However, our framework for periphery plots includes a set of user interactions that build upon the core concepts defined in this section to help support interactive data exploration.

\textbf{Zooming and Panning.} 
Users can zoom and pan through the overall time scale to explore data. Users can execute these interactions from the focus zone on any track, using the scroll-wheel to zoom or dragging the mouse to pan. These interactions are not supported in periphery plots, which may have a drastically different time scale (or none at all). Similar zoom and pan mouse gestures can be performed via the control timeline. Regardless of where a pan or zoom originates, all tracks are kept synchronized to maintain temporal alignment of the various focus and context zones.

Additionally, mouse hovers within any focus zone will display a vertical time indicator across all tracks. This helps facilitate navigation of the time axis, identification of specific date and time values, and visual alignment of data in the time axis.

\textbf{Multi-Brush Zone Control.} Beyond the aforementioned pan and zoom interactions, all other manipulations of the focus and context zones are performed via the control timeline. A pan or zoom event may cause the extent of a context zone to extend beyond the limits of the time axis. This allows users to move the focus zone to the very start or end of the overall time axis without having to remove or adjust context zones.

Users can change the extent of a single zone by manipulating the zone's corresponding brush on the control timeline. Each brush includes two \emph{handles} representing the start and end of the corresponding zone. The handles are visible as green squares in Figure~\ref{fig:teaser}(b$_1$,c,b$_2$). Users can drag any handle to resize the corresponding focus or control zone. Other zones will retain their size, but will be ``pushed'' or ``pulled'' to maintain the requirement that all zones are contiguous without gaps between them. 

The ``push'' and ``pull'' behavior is not always desired. For this reason, users can enable \emph{locks} to anchor a zone boundary point in time. The locks, visible in Figure~\ref{fig:teaser}(b$_1$,c,b$_2$) as black squares, can be enabled or disabled by clicking. When a lock is enabled, the handles above it are disabled since the corresponding boundary cannot be modified until the lock is removed. Locks also impact panning behaviors. This allows, for example, a user to lock context zones at the very start and end of the control timeline to create periphery plots that summarize ``all data before'' and ``all data after'' the focus zone.

%% file: sections/prototype.tex
\begin{figure*}[t]
\centering
\includegraphics[width=6.65in]{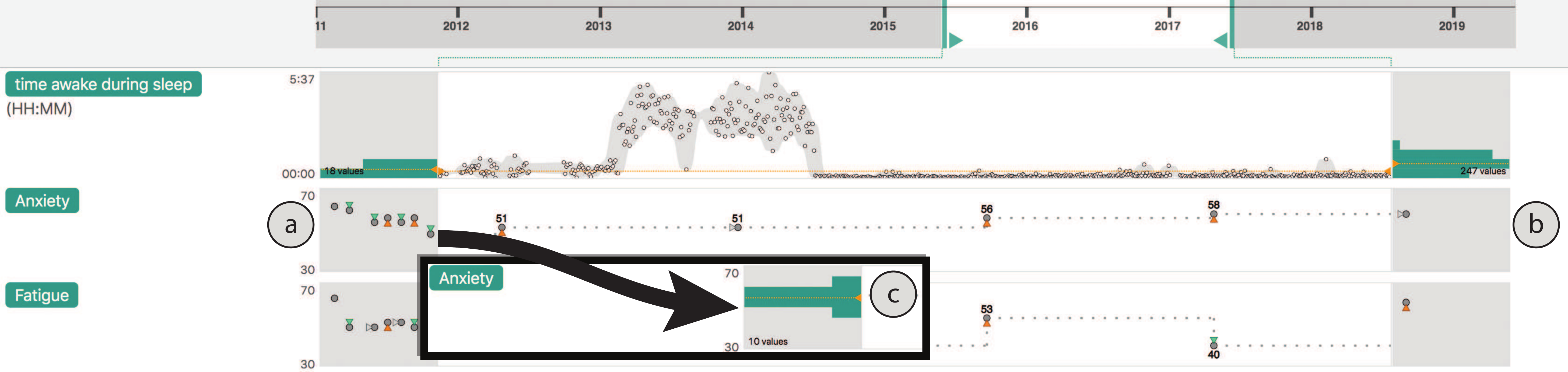}
\vspace{-0.19cm}
\caption{The Precision VISSTA App for data exploration uses periphery plots to help users understand how the values in the current focus area relate to data recorded prior to or after the focus time period. This example shows sparse and discrete values shown in two formats: (a,b) time and value preserving charts that condense the time axis, or (c) value-preserving histogram charts that eliminate the time axis in order to display data about a higher number of observations. Periphery plots can be configured to automatically transition between these modes.  
}
\vspace{-0.15cm}
\label{fig:health}
\end{figure*}

\section{Toolkit and Example Use Cases}
\label{sec:usecases}

Periphery plots were developed in the context of a broader project (Precision VISSTA \cite{qubbd_www}) focused on patient-generated health data (PGHD). The techniques were then generalized within a reusable toolkit for easy application to support the analysis of other datasets and domains. This section describes the reusable periphery plot toolkit and two example applications.

\subsection{Toolkit}
\label{sec:toolkit}

The core components of the periphery plot framework were first developed in JavaScript as part of the patient-generated health data platform described in Section~\ref{sec:health}. That application leverages the React framework and D3.js \cite{Bostock2011-oi} to implement the control timeline, individual data tracks, and the corresponding periphery plots. To facilitate reuse, we extracted these components and bundled them as a generic periphery plot toolkit. The toolkit provides all core capabilities within a React-based component that can be: (1) easily integrated into React-based web applications; and (2) easily customized and extended with new graphic styles, chart types, annotations, or interactions. The source code has been released as open source software~\cite{github_repo}.

\subsection{Use Case 1: Patient-Generated Health Data}
\label{sec:health}

Periphery plots were first developed to support visualizations of PGHD for clinical interpretation~\cite{qubbd_www}. This includes mHealth data gathered from various sensors (e.g., total steps or time awake during sleep) and manually entered logs (e.g., nutrition), as well as patient-reported outcomes data gathered via validated electronic surveys. PGHD contains 
multiple longitudinal variables with heterogeneous data types. Moreover, these variables are recorded at different rates with varied sparseness, and exhibit different forms of missingness. To make this complex combination of longitudinal records clinically useful, a visualization must allow users to quickly look for long term trends, outliers, patterns, and correlations across variables.

We applied the periphery plot framework to this problem as shown in Figure~\ref{fig:health}. The vertical alignment across tracks allows comparisons across variables. The most widely used periphery plot chart type was a value-preserving histogram showing the distribution of values in the pre- and post-focus context periods. This helps to show, for example, that the spikes in the \emph{time awake during sleep} track are unusually high compared to values in the context zones (which are set to show all data prior to or after the focus).

When the data in a context region is sufficiently sparse, a value-and-time preserving periphery plot chart type is used. This was motivated by the relatively sparse nature of many of the PGHD variables which are collected manually (rather than via sensors). Figure~\ref{fig:health}(a) exploits this feature to show not only that earlier anxiety scores were higher, but that they have been decreasing leading up to the focus time period. This feature is very useful, but does not scale well as the number of values in the context increases. For this reason, periphery plots are designed to automatically transition from (a)~the value-and-time preserving view to (c)~a value-preserving summarization, if a user interaction results in a higher number of samples in the periphery that exceeds a defined threshold.

\subsection{Use Case 2: Climate Data}

To demonstrate the versatility of periphery plots, we applied the toolkit to a second data set: a publicly available collection of Seattle weather observations \cite{seattle_www}. This use case is highlighted in Figure~\ref{fig:teaser}.

Compared to the health application above, weather data is more homogeneous (mostly scalars over time) and is more uniformly sampled over time. Moreover, it exhibits less sparsity and missingness. Yet despite these differences, the periphery plot framework can be easily applied to this use case. This application was developed with minimal effort by connecting CSV data with built-in chart types.

The interactive application allows for the discovery of several interesting observations. The screenshot in Figure~\ref{fig:teaser} shows examples, including the seasonal fluctuations in temperatures highlighted in the \emph{temp max} track. Here, the focus and context zones were each set to approximately six months in duration. The annotations in that track, Figure~\ref{fig:teaser}(e), show that the focus period is much hotter than the period before and after. Panning the control timeline forward or back shows this pattern oscillate as the focus moves from winter to summer an back again. 
A second finding is shown in Figure~\ref{fig:teaser}(d$_4$) and highlights some drizzling precipitation. This is quite different from the current focus, which is more similar to the distribution of observations shown in the post-focus context region in Figure~\ref{fig:teaser}(d$_8$).

%% file: sections/conclusion.tex
\section{Discussion and Conclusion}

The use cases in Section~\ref{sec:usecases} illustrate the strengths of periphery plots in contextualizing detailed time-based data within a larger time frame. The examples demonstrate, in particular, how the periphery plot framework provides a flexible approach for visual encoding in the context zone. This includes the automated transition between plot types to adjust the level of information preservation based on available space. There are, however, some limitations in our current definition of the framework. For example, the framework does not explicitly cover a scenario in which multiple variables are shown within the same track. 
Furthermore, the framework does not ensure that periphery plots are easy to compare within or between tracks.  In future work, we hope to explore solutions to this challenge by building upon prior research into the design space and best practices for multi-view comparability and consistency \cite{qu_keeping_2018,knudsen_view_2016}.

In summary, we demonstrate that the periphery plot framework for providing contextual information for heterogeneous time-based data is generalizable and delivers insights that would not be possible without the context provided by our approach. Despite the relatively limited design space, periphery plots are an effective approach that is easy to implement and use. Furthermore, the approach does not require any complex visual representations or interactions to convey information critical to the interpretation of the data. Therefore, we anticipate that periphery plots have the potential to be effective for a wide range of users across multiple application domains.